\documentclass[12pt]{article}
\usepackage{mathrsfs}
\usepackage{amsfonts}
\usepackage{amsmath}
\usepackage{amssymb}
\usepackage{axodraw}
\usepackage{mathptmx}
\usepackage[dvips]{color}
\usepackage{epsfig}
\usepackage{graphicx}
\newcommand{\calA}{{\cal A}}

\newcommand{\kslash}{/\!\!\!\!\!k}
\newcommand{\pslash}{/\!\!\!\!\!p}
\newcommand{\qslash}{/\!\!\!\!\!q}

\begin{document}
\baselineskip=16pt

\pagenumbering{arabic}

\vspace{1.0cm}

\begin{center}
{\Large\sf Gauge independence of magnetic moment and vanishing
charge of Dirac neutrinos: an exact one-loop demonstration}
\\[10pt]
\vspace{.5 cm}

{Wen-Tao Hou$^{a,b}$, Yi Liao$^a$\footnote{liaoy@nankai.edu.cn}, and Hong-Jun Liu$^a$}

{\it $^a$ School of Physics, Nankai University, Tianjin 300071, China
\\
$^b$ Department of Modern Physics, University of Science and
Technology of China,
\\
Hefei 230026, Anhui, China}

\vspace{2.0ex}

{\bf Abstract}

\end{center}

The magnetic moment and vanishing charge of a Dirac neutrino are
physically observable quantities and must not depend on the choice
of gauge in a consistent quantum field theory. We verify this
statement explicitly at the one loop level in both $R_\xi$ and
unitary gauges of the minimally extended standard model. We
accomplish this by manipulating directly the integrands of loop
integrals and employing simple algebraic identities and integral
relations. Our result generally applies for any masses of the
relevant particles and unitary neutrino mixing.

\begin{flushleft}
PACS: 14.60.Lm, 13.40.Em, 12.15.Lk


\end{flushleft}

\newpage

\section{Introduction}
\label{sec:intro}

The fundamental properties of a particle like its charge and
electromagnetic dipole moments are physical quantities that in
principle are experimentally measurable, for reviews, see Refs.
\cite{Pospelov:2005pr,Giunti:2008ve}. These quantities can be
unambiguously calculated in a quantum-mechanically consistent theory
like the standard model (SM), and confronted with the measurements
to decide whether the theory is correct or not. That said, the
practical calculation and demonstration of its result being
independent of computational methods are not always trivial. We have
witnessed a similar circumstance recently, concerning the one-loop
contribution of the charged weak gauge bosons $W^\pm$ to the
two-photon decay rate of the Higgs boson. A new computation in
unitary gauge \cite{Gastmans:2011ks,Gastmans:2011wh} claimed an
answer that is different from the well-spread result obtained long
ago
\cite{Ellis:1975ap,Ioffe:1976sd,Shifman:1979eb,Vainshtein:1980ea} in
a special gauge, i.e., the 't Hooft-Feynman gauge ($\xi=1$) among
the class of renormalizable $R_\xi$ gauges. Subsequent studies by
various methods, including computing in both $R_\xi$ and unitary
gauges, see for instance Refs. \cite{Marciano:2011gm,Shao:2011wx},
confirmed the old result, and taught us a great deal on
computational subtleties in a theory that is nontrivial in the high
energy regime.

In this work we examine a similar problem in the neutrino sector,
i.e., the charge and magnetic moment of a Dirac neutrino in SM that
is minimally extended by the introduction of right-handed neutrinos.
We show explicitly at the one-loop level in both $R_\xi$ and unitary
gauges that the neutrino charge vanishes and its magnetic moment is
a gauge independent quantity. The issue has been partially studied
in the literature. Early works
\cite{Bardeen:1972vi,Lee:1977tib,Beg:1977xz,Marciano:1977wx,Fujikawa:1980yx}
assumed a massless neutrino or expanded the quantities to the
leading order in the small masses of neutrinos and charged leptons,
ignored the lepton mixing, or computed in a special gauge. A further
step was taken some years ago
\cite{CabralRosetti:1999ad,Dvornikov:2003js,Dvornikov:2004sj}. It
was found \cite{Dvornikov:2003js}, for instance, that up to the
second order in the expansion of small neutrino masses the charge
vanishes and the magnetic moment is $\xi$-independent, and that the
charge vanishes exactly in the 't Hooft-Feynman gauge. Here we cope
directly with the integrands of loop integrals, and demonstrate
manifestly that both quantities are gauge independent for any masses
of the relevant particles and for any unitary lepton mixing.

In the next section we set up our notations and suggest how to
calculate in a nice way to isolate the terms that potentially
contribute to the charge and magnetic moment. We describe in some
detail in sec \ref{sec:Rgau} our calculation in $R_\xi$ gauge. Our
one-loop exact result for the magnetic form factor at vanishing
momentum transfer is shown in Eq.(\ref{eq_F20}). This is followed by
a short discussion in sec \ref{sec:Ugau} on the calculation in
unitary gauge. We summarize briefly in the last section.

\section{Computational strategy}
\label{sec:pre}

The charge and magnetic moment of a Dirac particle can be defined by
the amplitude of a process in which it radiates a photon,
\begin{eqnarray}
\bar u(p_-)i\calA_\mu(q)u(p_+)=(-ie)\bar u(p_-)\Big[\gamma_\mu
F_1(q^2)-\frac{1}{2m}i\sigma_{\mu\nu}q^\nu F_2(q^2)
+\cdots\Big]u(p_+).
\label{eq_A}%
\end{eqnarray}
Here $p_\pm=p\pm q/2$ are the momenta of the incoming and outgoing
particle of mass $m$, and $q$ is the photon's outgoing momentum. The
above decomposition in terms of the standard form factors is based
on Lorentz covariance and electromagnetic gauge invariance, and
assumes that the Dirac particle in both initial and final states is
physical:
\begin{eqnarray}
\pslash_\pm u(p_\pm)=mu(p_\pm),~p_\pm^2=m^2.
\label{eq_EoM}%
\end{eqnarray}
The dots in Eq. (\ref{eq_A}) stand for two more form factors that
are irrelevant here; one corresponds to the electric dipole moment
that cannot occur at one loop in the minimally extended SM (as can
also be seen from sec \ref{sec:Rgau}), and the other is the
so-called anapole whose Lorentz structure is quadratic in $q$. The
form factors at an arbitrary $q^2$ are generally not measurable
quantities, since the above (unphysical) amplitude appears as part
of the complete contribution to a physical process. Nevertheless,
$F_1(0)$ and $F_2(0)$ are physical quantities because they
correspond to the charge and anomalous magnetic moment of the
particle. Our convention is such that the electron has the charge
$eF_1(0)=e<0$ and the magnetic moment vector,
$\vec\mu=(e/m)[F_1(0)+F_2(0)]\vec S$ with $\vec S$ being its spin
vector, that appears, e.g., in the interaction potential of the
dipole with an external magnetic field $\vec B$,
$V=-\vec\mu\cdot\vec B$.

The charge, $F_1(0)$, is relatively easy to isolate. Setting $q=0$
removes all other Lorentz structures, and allows us to employ the
equations of motion (EoMs) in the limit $q\to 0$ for both initial
and final particles, $\pslash u(p)=mu(p)$, to reduce the amplitude
completely to the $\gamma_\mu$ form. There are several ways to work
out the anomalous magnetic moment $F_2(0)$. One could isolate by
brute force terms contributing to the form factor $F_2(q^2)$ and
take its value at $q^2=0$. Most studies in the literature follow
this approach. In the second approach, one employs a projection
operator, and expresses $F_2(0)$ as a combination of Dirac traces
\cite{Roskies:1990ki,Czarnecki:1996if,Czarnecki:1996rx}. Here we
take a third approach, which might be the best to observe the
cancellation of gauge dependence among various Feynman graphs. As we
will show in the next section, the cancellation happens at the level
of loop integrands. In this approach, we take the derivative of the
amplitude with respect to the photon momentum,
$i\partial_\nu^q\calA_\mu(q)$, antisymmetrize it in the Lorentz
indices $\mu$ and $\nu$, and then evaluate it at $q=0$. Since all
form factors are smooth at $q^2=0$, only the magnetic moment term
survives the procedure and yields $-e/(2m)\sigma_{\mu\nu}F_2(0)$.
[We remind once again that the electric dipole term vanishes at one
loop but would appear at higher orders.] Comparison of the two gives
the answer for $F_2(0)$.

An important point in implementing the above procedure should be
noted. We mentioned that the decomposition in eq (\ref{eq_A}) is
possible only upon using EoMs (\ref{eq_EoM}). When computing
$F_2(0)$, we are essentially expanding $\calA_\mu(q)$ in small $q$
and isolating its linear terms. A term that is manifestly linear in
$q$ cannot avoid our eyes, for which we are free to apply the
limiting EoMs, $\pslash u(p)=mu(p)$, because the difference to the
exact ones does not affect $F_2(0)$. With terms of apparently zeroth
order in $q$ we should be careful. For these terms, when necessary,
we must apply the exact equations (\ref{eq_EoM}) since the
difference now is exactly what we are interested in and may enter
$F_2(0)$. Ignoring this will result in an incorrect, gauge-dependent
answer. Another point is more technical. Although antisymmetrization
in Lorentz indices is not mandatory since it will come out
automatically upon finishing the calculation, one can simplify the
algebra by doing antisymmtrization at an early stage.

\begin{center}
\begin{picture}(400,180)(0,0)

\SetOffset(10,120)
\ArrowLine(0,0)(20,0)\ArrowLine(20,0)(80,0)\ArrowLine(80,0)(100,0)%
\Photon(20,0)(50,40){3}{7}\Photon(80,0)(50,40){-3}{7}%
\Photon(50,40)(50,55){-3}{2.5}%
\Text(0,-8)[r]{$\nu_i$}\Text(50,-8)[]{$\ell_\alpha$}
\Text(100,-8)[l]{$\nu_i$}%
\Text(10,6)[]{$p_+$}\Text(50,6)[]{$k\!\!+\!\!p$}\Text(90,6)[l]{$p_-$}%
\Text(55,50)[l]{$q\!\uparrow$}\Text(50,60)[]{$\gamma~_\mu$}%
\Text(22,25)[]{$W^-$}\Text(82,25)[]{$W^-$}%
\Text(38,18)[l]{$k_-$}\Text(65,18)[r]{$k_+$}%
\Text(22,15)[]{$_\rho$}\Text(82,15)[]{$_\sigma$}
\Text(35,35)[]{$_\alpha$}\Text(65,35)[]{$_\beta$}
\Text(50,-25)[]{$(a)$}

\SetOffset(140,120)
\ArrowLine(0,0)(20,0)\ArrowLine(20,0)(80,0)\ArrowLine(80,0)(100,0)%
\DashLine(20,0)(50,40){3}\DashLine(80,0)(50,40){3}%
\Photon(50,40)(50,55){-3}{2.5}%
\Text(22,25)[]{$G^-$}\Text(82,25)[]{$G^-$}%
\Text(50,-25)[]{$(b)$}

\SetOffset(270,120)
\ArrowLine(0,0)(20,0)\ArrowLine(20,0)(80,0)\ArrowLine(80,0)(100,0)%
\DashLine(20,0)(50,40){3}\Photon(80,0)(50,40){-3}{7}%
\Photon(50,40)(50,55){-3}{2.5}%
\Text(50,-25)[]{$(c)$}

\SetOffset(10,20)
\ArrowLine(0,0)(20,0)\ArrowLine(20,0)(80,0)\ArrowLine(80,0)(100,0)%
\DashLine(80,0)(50,40){3}\Photon(20,0)(50,40){3}{7}%
\Photon(50,40)(50,55){-3}{2.5}%
\Text(50,-25)[]{$(d)$}

\SetOffset(140,20)
\ArrowLine(0,30)(20,30)\ArrowLine(20,30)(50,30)\ArrowLine(50,30)(80,30)
\ArrowLine(80,30)(100,30)%
\PhotonArc(50,30)(30,180,360){3}{10}
\Photon(50,30)(50,55){-3}{3}%
\Text(10,36)[]{$p_+$}\Text(33,36)[]{$k\!\!+\!\!p_+$}
\Text(67,36)[]{$k\!\!+\!\!p_-$}\Text(90,36)[l]{$p_-$}%
\Text(50,-20)[]{$(e)$}

\SetOffset(270,20)
\ArrowLine(0,30)(20,30)\ArrowLine(20,30)(50,30)\ArrowLine(50,30)(80,30)
\ArrowLine(80,30)(100,30)%
\DashCArc(50,30)(30,180,360){3}%
\Photon(50,30)(50,55){-3}{3}%
\Text(50,-20)[]{$(f)$}

\end{picture}
\end{center}

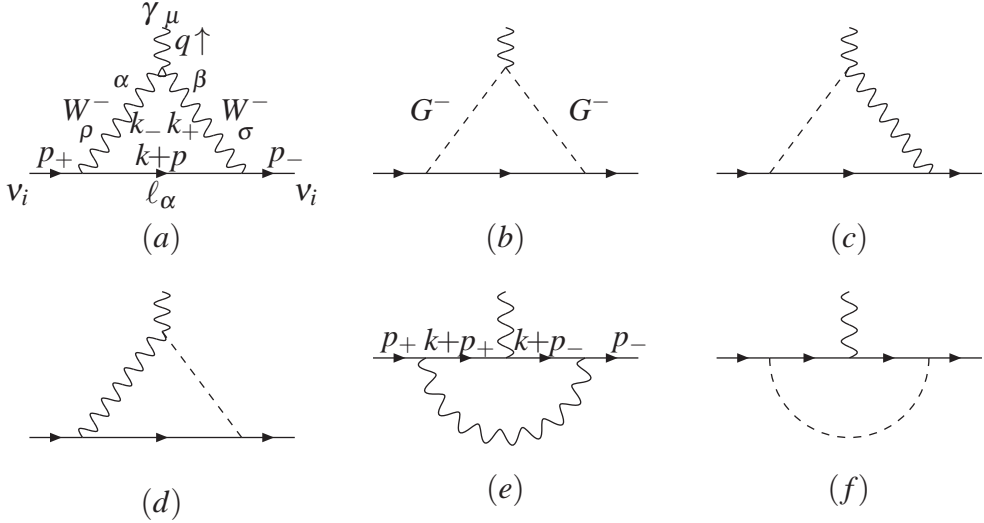
Fig. 1 Feynman diagrams contributing at one loop to the vertex
function $i\Gamma_\mu(q)$. Wavy (dashed, dotted, solid) lines stand
for the gauge boson (scalar, ghost, fermion) fields.

\begin{center}
\begin{picture}(350,170)(0,0)

\SetOffset(10,120)%
\Photon(30,5)(0,0){2}{4}\Photon(30,5)(60,0){2}{4}\PhotonArc(30,25)(20,0,360){2}{18}
\Text(0,-8)[r]{$A_\mu$}\Text(60,-8)[l]{$Z_\nu$}
\Text(10,10)[r]{$_\rho$}\Text(50,10)[l]{$_\sigma$}\Text(55,25)[l]{$W^-$}%
\Text(30,-25)[]{$(a)$}

\SetOffset(90,120)
\Photon(30,5)(0,0){2}{4}\Photon(30,5)(60,0){2}{4}\DashCArc(30,25)(20,0,360){3}
\Text(55,25)[l]{$G^-$}%
\Text(30,-25)[]{$(b)$}

\SetOffset(170,120)%
\Photon(15,20)(0,20){2}{2.5}\Photon(55,20)(70,20){-2}{2.5}
\PhotonArc(35,20)(20,0,360){2}{18}%
\Text(35,30)[]{$W^-$}\Text(15,10)[r]{$_\rho$}\Text(15,30)[r]{$_\alpha$}%
\Text(55,10)[l]{$_\sigma$}\Text(55,30)[l]{$_\beta$}%
\Text(35,-25)[]{$(c)$}

\SetOffset(250,120)%
\Photon(15,20)(0,20){2}{2.5}\Photon(55,20)(70,20){-2}{2.5}
\DashCArc(35,20)(20,0,360){3}\Text(35,30)[]{$G^-$}%
\Text(35,-25)[]{$(d)$}

\SetOffset(10,30)%
\Photon(15,20)(0,20){2}{2.5}\Photon(55,20)(70,20){-2}{2.5}
\PhotonArc(35,20)(20,0,180){2}{9}\DashCArc(35,20)(20,180,360){3}%
\Text(35,30)[]{$W^\mp$}\Text(35,10)[]{$G^\mp$}%
\Text(15,30)[r]{$_\alpha$}\Text(55,30)[l]{$_\beta$}
\Text(35,-25)[]{$(e)$}

\SetOffset(90,30)%
\Photon(15,20)(0,20){2}{2.5}\Photon(55,20)(70,20){-2}{2.5}
\DashArrowArc(35,20)(20,0,180){1.5}\DashArrowArc(35,20)(20,180,360){1.5}%
\Text(35,30)[]{$c^\pm$}%
\Text(35,-25)[]{$(f)$}

\SetOffset(170,30)%
\Photon(15,20)(0,20){2}{2.5}\Photon(55,20)(70,20){-2}{2.5}
\ArrowArc(35,20)(20,0,180)\ArrowArc(35,20)(20,180,360)%
\Text(35,30)[]{$f$}%
\Text(35,-25)[]{$(g)$}

\SetOffset(250,30)%
\Photon(0,20)(15,20){2}{3}\BCirc(25,20){10}\Photon(35,20)(50,20){2}{3}
\ArrowLine(60,0)(50,20)\ArrowLine(50,20)(60,40)

\end{picture}
\end{center}

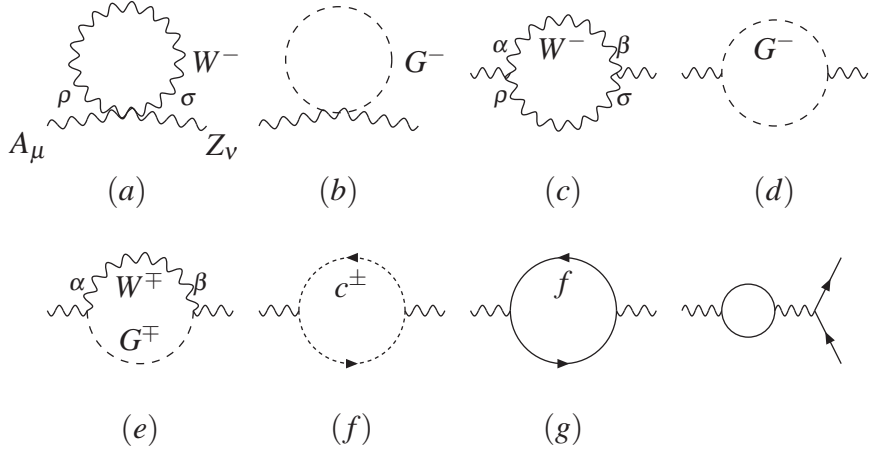
Fig. 2 Feynman diagrams contributing at one loop to the $\gamma Z$
mixing energy $i\Pi_{\mu\nu}(q)$.

There are two classes of Feynman graphs in SM that contribute at one
loop to the amplitude $i\calA_\mu(q)$, through the proper vertex
$i\Gamma_\mu(q)$ in Fig. 1 and the photon-$Z$ boson mixing energy in
Fig. 2, $i\Pi_{\mu\nu}(q)$, attached to the tree level neutrino-$Z$
vertex (see the last graph in Fig. 2). While the former contributes
to both $F_1(0)$ and $F_2(0)$,  the latter contributes only to
$F_1(0)$ through
\begin{eqnarray}
i\Pi_{\mu\nu}(0)\frac{i}{m_Z^2}\frac{ig_2}{2c_W}\gamma^\nu P_L.
\label{eq_mixing}%
\end{eqnarray}
Here we use the standard notations of SM: $m_{W,Z}$ are the masses
of the $W^\pm$ and $Z$ bosons, $g_2$ is the gauge coupling of
$SU(2)_L$, $c_W=\cos\theta_W$ and $s_W=\sin\theta_W$ with $\theta_W$
being the weak mixing angle, and $P_{L,R}=(1\mp\gamma_5)/2$. We
display here the contributions from individual graphs. Working in
$d$-dimensions, we write
\begin{eqnarray}
i\Gamma_\mu(q)&=&\frac{1}{2}eg_2^2|V_{\alpha i}|^2\sum_{x=a}^f\int_k(1x),%
\label{eq_Gamma}%
\\
i\Pi_{\mu\nu}(0)&=&\frac{eg_2}{c_W}\sum_{x=a}^f
\int_k(2x),~\int_k\equiv\int\frac{d^dk}{(2\pi)^d},
\label{eq_mixingenergy}%
\end{eqnarray}
where, denoting $k_\pm=k\pm q/2$, from Fig. 1,
\begin{eqnarray}
(1a)&=&+\gamma_\sigma P_L(\kslash+\pslash+m_\alpha)\gamma_\rho P_L
\Gamma_{\alpha\beta\mu}(-k_-,k_+,-q)P^{\alpha\rho}(k_-)P^{\beta\sigma}(k_+)P^{-1},
\\
(1b)&=&+m_W^{-2}\frac{(m_iP_L-m_\alpha
P_R)(\kslash+\pslash+m_\alpha)(m_iP_R-m_\alpha P_L)(k_-+k_+)_\mu}
{(k_+^2-\xi_W m_W^2)(k_-^2-\xi_W m_W^2)P},
\\
(1c)&=&+\frac{\gamma_\sigma P_L
(\kslash+\pslash+m_\alpha)(m_iP_R-m_\alpha P_L)
P^{\mu\sigma}(k_+)}{[k_-^2-\xi_W m_W^2]P},
\\
(1d)&=&+\frac{(m_i P_L-m_\alpha P_R)(\kslash+\pslash+m_\alpha)
\gamma_\rho P_LP^{\mu\rho}(k_-)}{[k_+^2-\xi_W m_W^2]P},
\\
(1e)&=&-\frac{\gamma_\sigma
P_L(\kslash+\pslash_-+m_\alpha)\gamma_\mu
(\kslash+\pslash_++m_\alpha)\gamma_\rho P_LP^{\rho\sigma}(k)}
{[(k+p_+)^2-m_\alpha^2][(k+p_-)^2-m_\alpha^2]},
\\
(1f)&=&+m_W^{-2}\frac{(m_iP_L-m_\alpha P_R)
(\kslash+\pslash_-+m_\alpha)\gamma_\mu(\kslash+\pslash_++m_\alpha)(m_iP_R-m_\alpha
P_L)} {[(k+p_+)^2-m_\alpha^2][(k+p_-)^2-m_\alpha^2]Q_2},
\end{eqnarray}
and from Fig. 2,
\begin{eqnarray}
(2a)&=&-c_W^2[2g_{\rho\sigma}g_{\mu\nu}-g_{\rho\mu}g_{\sigma\nu}
-g_{\rho\nu}g_{\sigma\mu}]P^{\rho\sigma},
\\
(2b)&=&-(c_W^2-s_W^2)g_{\mu\nu}Q_2^{-1},
\\
(2c)&=&+c_W^2\Gamma_{\rho\alpha\mu}(-k,k,0)
\Gamma_{\beta\sigma\nu}(-k,k,0)P^{\alpha\beta}P^{\rho\sigma},
\\
(2d)&=&+2(c_W^2-s_W^2)k_\mu k_\nu Q_2^{-2},
\\
(2e)&=&+2s_W^2m_W^2P_{\mu\nu}Q_2^{-1},
\\
(2f)&=&-2c_W^2k_\mu k_\nu Q_2^{-2}.
\end{eqnarray}
Note that the fermion loop in Fig. 2(g) is transverse and drops out
at $q=0$. We have defined the shortcuts for the propagators and
triple-gauge vertex:
\begin{eqnarray}
&&\Gamma_{\alpha\beta\mu}(p_1,p_2,p_3)=(p_2-p_3)_\alpha g_{\beta\mu}
+(p_3-p_1)_\beta g_{\mu\alpha}+(p_1-p_2)_\mu g_{\alpha\beta},
\label{eq_triple}
\\
&&P_{\mu\nu}(p)=g_{\mu\nu}[p^2-m_W^2]^{-1}-\delta_Wp_\mu
p_\nu[p^2-\xi_Wm_W^2]^{-1}[p^2-m_W^2]^{-1},
\label{eq_PropW}%
\\
&&P=(k+p)^2-m_\alpha^2,~Q_1=k^2-m_W^2,~Q_2=k^2-\xi_Wm_W^2,
\label{eq_prop}%
\end{eqnarray}
with $P_{\alpha\beta}=P_{\alpha\beta}(k)$ and $\delta_W=1-\xi_W$.
$m_\alpha$ ($m_i$) is the mass of the charged lepton $\ell_\alpha$
(neutrino $\nu_i$), and $V_{\alpha i}$ is the lepton mixing matrix
appearing in the charged current interaction. A summation over all
$\ell_\alpha$ is always implied. The identical initial and final
neutrino satisfies EoMs (\ref{eq_EoM}) where now $m=m_i$. The above
loop integrands will be manipulated in the next two sections.

\section{Evaluation in $R_\xi$ gauge}
\label{sec:Rgau}

\subsection{Charge}

Let us start with the charge. Setting $q=0$ simplifies significantly
the expressions of $(1x)$. Using
$(\kslash+\pslash)\gamma_\mu(\kslash+\pslash)=-(k+p)^2\gamma_\mu+2(k+p)_\mu(\kslash+\pslash)$,
$\partial_\mu Q_2^{-1}=-2k_\mu Q_2^{-2}$, and $\partial_\mu
P^{-1}=-2(k+p)_\mu P^{-2}$, $(1b)$ and $(1f)$ sum to a total
derivative:
\begin{eqnarray}
[(1b)+(1f)]_0&=&-m_W^{-2}\partial_\mu\big\{\big[(\kslash+\pslash)(m_i^2P_R+m_\alpha^2 P_L)%
-m_im_\alpha^2\big](PQ_2)^{-1}\big\},
\end{eqnarray}
where the subscript $0$ denotes evaluation at $q=0$. Considering the
relation
\begin{eqnarray}
\Gamma_{\alpha\beta\mu}(-k,k,0)P^{\alpha\rho}P^{\beta\sigma}
=(k^\rho P^\sigma_\mu+k^\sigma P^\rho_\mu)Q_2^{-1}+
\partial_\mu P^{\rho\sigma},
\end{eqnarray}
we combine the pure $W^\pm$-loop graphs,
\begin{eqnarray}
[(1a)+(1e)]_0&=&+\partial^\mu\big\{\gamma_\sigma(\kslash+\pslash)\gamma_\rho
P_LP^{-1}P^{\rho\sigma}\big\}%
\nonumber
\\
&&+P_R\gamma_\sigma(\kslash+\pslash)\kslash P^{\sigma\mu}(PQ_2)^{-1}%
+\kslash(\kslash+\pslash)\gamma_\rho P_LP^{\rho\mu}(PQ_2)^{-1}.
\label{eq_1aplus1e0}%
\end{eqnarray}
The last two terms in the above are summed with the remaining two
graphs to yield
\begin{eqnarray}
[(1a)+(1e)+(1c)+(1d)]_0&=&+\partial^\mu\big\{\gamma_\sigma(\kslash+\pslash)\gamma_\rho
P_LP^{-1}P^{\rho\sigma}\big\}%
\nonumber
\\
&&+P_R\gamma_\sigma
\big[(\kslash+\pslash)(\kslash+m_i)-m_\alpha^2)\big]P^{\mu\sigma}(PQ_2)^{-1}
\nonumber
\\
&&+\big[(\kslash+m_i)(\kslash+\pslash)-m_\alpha^2\big]\gamma_\rho
P_LP^{\mu\rho}(PQ_2)^{-1}.
\end{eqnarray}
Since the above expression is sandwiched between the spinors of the
initial and final states, it is tempting to replace $m_i$ in the
last two terms by $\pslash$. But this is not legitimate as
emphasized in the last section. Instead, $m_i$ should be replaced by
$(\pslash\pm\qslash/2)$ on the rightmost (leftmost), in terms of the
exact EoMs (\ref{eq_EoM}):
\begin{eqnarray}
&&\bar u(p_-)[(1a)+(1e)+(1c)+(1d)]_0u(p_+)
\nonumber
\\
&=&\bar u(p_-)\big(
\partial^\mu\big\{\gamma_\sigma(\kslash+\pslash)\gamma_\rho
P_LP^{-1}P^{\rho\sigma}\big\}%
+2\gamma_\rho P_LP^{\mu\rho}Q_2^{-1}\big)u(p_+)%
\nonumber
\\
&&+\bar u(p_-)(1cd)_qu(p_+),
\end{eqnarray}
where the last term linear in $q$ does not contribute to the charge
but may contribute to the magnetic moment,
\begin{eqnarray}
(1cd)_q&=&\frac{1}{2}P_R\gamma_\sigma(\kslash+\pslash)\qslash
P^{\mu\sigma}(PQ_2)^{-1}%
-\frac{1}{2}\qslash(\kslash+\pslash)\gamma_\rho
P_LP^{\mu\rho}(PQ_2)^{-1}.
\label{eq_1cd}%
\end{eqnarray}
In summary, leaving aside the $(1cd)_q$ term, we have
\begin{eqnarray}
\sum_{x=a}^f(1x)_0&=&+\partial_\mu\big\{%
-m_W^{-2}\big[(\kslash+\pslash)(m_i^2P_R+m_\alpha^2P_L)-m_im_\alpha^2\big](PQ_2)^{-1}%
\nonumber
\\
&&+\gamma_\sigma(\kslash+\pslash)\gamma_\rho P_LP^{\rho\sigma}P^{-1}\big\}%
+2\gamma_\rho P_LP_\mu^\rho Q_2^{-1}.
\end{eqnarray}
The total derivative can be dropped in regularized loop integrals,
so that Fig. 1 contributes to the $F_1(0)$ term in Eq. (\ref{eq_A})
the following:
\begin{eqnarray}
+eg_2^2\sum_\alpha|V_{\alpha i}|^2
\int_k\gamma_\rho P_LP_\mu^\rho Q_2^{-1}%
=+eg_2^2\int_k\gamma_\rho P_LP_\mu^\rho Q_2^{-1},
\label{eq_charge1}%
\end{eqnarray}
where unitarity of $V$ is used to finish the sum as the integrand is
independent of $m_\alpha$.

Now we manipulate $i\Pi_{\mu\nu}(0)$. First of all, $(2b)$ and
$(2d)$ form a total derivative:
\begin{eqnarray}
(2b)+(2d)&=&-(c_W^2-s_W^2)\partial_\mu\big(k_\nu Q_2^{-1}\big).
\end{eqnarray}
Using the shortcuts in Eqs. (\ref{eq_triple},\ref{eq_PropW}), we
have
\begin{eqnarray}
(2a)&=&2c_W^2\big[\delta_\xi(k^2g_{\mu\nu}-k_\mu k_\nu)(Q_1Q_2)^{-1}
-g_{\mu\nu}(d-1)Q_1^{-1}\big],
\label{eq_2a}%
\\
(2c)&=&2c_W^2\big[\xi_W(k^2g_{\mu\nu}-k_\mu k_\nu)(Q_1Q_2)^{-1}
+2(d-1)k_\mu k_\nu(Q_1)^{-2}\big].
\label{eq_2c}%
\end{eqnarray}
The last terms in $(2a)$ and $(2c)$ already form a total derivative.
In the first terms, we decompose
$k^2(Q_1Q_2)^{-1}=Q_2^{-1}+m_W^2(Q_1Q_2)^{-1}$, and then sum
judiciously with $(2f)$ to arrive at the result
\begin{eqnarray}
(2a)+(2c)+(2f)&=&2c_W^2\Big\{-(d-1)\partial_\mu(k_\nu Q_1^{-1})
+g_{\mu\nu}\big[Q_2^{-1}+m_W^2(Q_1Q_2)^{-1}\big]
\nonumber%
\\
&&+k_\mu k_\nu\big[Q_2^{-2}-(Q_1Q_2)^{-1}\big]-2k_\mu k_\nu
Q_2^{-2}\Big\}
\nonumber%
\\
&=&2c_W^2\Big\{\partial_\mu\big[k_\nu Q_2^{-1}-(d-1)(k_\nu
Q_1^{-1})\big] +m_W^2Q_2^{-1}P_{\mu\nu}\Big\}.
\end{eqnarray}
Thus, using $c_W^2+s_W^2=1$, the sum of all graphs is
\begin{eqnarray}
\sum_{x=a}^f(2x)&=&\partial_\mu\big\{k_\nu
Q_2^{-1}-2c_W^2(d-1)(k_\nu Q_1^{-1})\big\}
+2m_W^2Q_2^{-1}P_{\mu\nu}.
\label{eq_sum2x}%
\end{eqnarray}
Dropping the regularized total derivative and using Eqs.
(\ref{eq_mixing}, \ref{eq_mixingenergy}), its contribution to the
$F_1(0)$ term in Eq. (\ref{eq_A}) is as follows,
\begin{eqnarray}
-eg_2^2\int_kQ_2^{-1}P_{\mu\nu}\gamma^\nu P_L, \label{eq_charge2}
\end{eqnarray}
which cancels Eq. (\ref{eq_charge1}). The vanishing charge is thus
established at one loop in $R_\xi$ gauge.

\subsection{Magnetic moment}
\label{subsec:mag}%

Moving to the magnetic moment, we follow the computational procedure
proposed in sec \ref{sec:pre}. Now only the graphs in Fig. 1
contribute. Since $(1b)$ is quadratic in $q$ when expanding in $q$,
it drops out. The next simplest is $(1f)$. Taking a derivative with
respect to $q^\nu$, setting $q=0$ and making it manifestly
antisymmetric in $\mu$ and $\nu$ (denoted by the pair of square
brackets below), we have
\begin{eqnarray}
\big[\partial_\nu^q(1f)_0\big]&=&+\frac{1}{4m_W^2}\frac{1}{P^2Q_2}\Big(
(K_{\mu\nu}^0+K_{\mu\nu}^1)(m_i^2P_R+m_\alpha^2P_L)%
-2m_im_\alpha^2[\gamma_\mu,\gamma_\nu]\Big),
\end{eqnarray}
where
\begin{eqnarray}
K_{\mu\nu}^0&=&\pslash[\gamma_\mu,\gamma_\nu]+[\gamma_\mu,\gamma_\nu]\pslash,
\\
K_{\mu\nu}^1&=&\kslash[\gamma_\mu,\gamma_\nu]+[\gamma_\mu,\gamma_\nu]\kslash.
\end{eqnarray}
Anticipating that $\big[\partial_\nu^q(1f)_0\big]$ is to be
sandwiched between the initial and final spinors and noting that the
$\kslash$ in $K_{\mu\nu}^1$ will yield a $\pslash$ upon loop
integration, we can apply the limiting EoMs $\pslash u=m_iu$ after
the moment has been isolated. The above is thus reduced to
\begin{eqnarray}
\big[\partial_\nu^q(1f)_0\big]&\leftrightharpoons&+\frac{1}{m_W^2}\frac{1}{8P^2Q_2}
\Big(K^1_{\mu\nu}(m_i^2+m_\alpha^2)+2m_i[\gamma_\mu,\gamma_\nu](m_i^2-m_\alpha^2)\Big),
\label{eq_dipole1f}%
\end{eqnarray}
where, from now on, $\leftrightharpoons$ means equality when
sandwiched between the spinors or under the loop integration or
both. All factors of $P_{L,R}$ are removed in a similar fashion,
confirming that the electric dipole moment does not arise at the one
loop.

Figs. $(1c)$ and $(1d)$ should be treated together for symmetry
reasons. There are two sources of terms, one from those explicitly
linear in $q$ and the other from the remaining terms (\ref{eq_1cd})
when computing the charge. Putting them together and taking the
derivative, we have
\begin{eqnarray}
\partial_\nu^q(1c+1d)_0&=&\big[m_i(\kslash+\pslash)\gamma^\sigma P_L
-P_R\gamma^\sigma m_i(\kslash+\pslash)\big]
\big[P_{\nu;\mu\sigma}Q_2^{-1}-k_\nu
Q_2^{-2}P_{\mu\sigma}\big]P^{-1}
\nonumber%
\\
&&+\frac{1}{2}\big[\gamma^\sigma(\kslash+\pslash)\gamma^\rho
-\gamma^\rho(\kslash+\pslash)\gamma^\sigma\big]P_L
g_{\rho\nu}P_{\mu\sigma}(PQ_2)^{-1},
\end{eqnarray}
where
\begin{eqnarray}
P_{\nu;\alpha\beta}&=&-\frac{1}{2}\partial_\nu P_{\alpha\beta}.
\end{eqnarray}
Antisymmetrization and applying EoMs yield, after some algebra,
\begin{eqnarray}
\big[\partial_\nu^q(1c+1d)_0\big]&\leftrightharpoons&m_iK_{\mu\nu}^2
\bigg(\frac{\delta_\xi}{4PQ_1Q_2^2}
-\frac{1}{4PQ_1^2Q_2}+\frac{1}{4PQ_1Q_2^2}\bigg)
\nonumber%
\\
&&
-\big(K_{\mu\nu}^1+2m_i[\gamma_\mu,\gamma_\nu]\big)\frac{1}{8PQ_1Q_2},
\label{eq_dipole1cd}%
\end{eqnarray}
where
\begin{eqnarray}
K^2_{\mu\nu}&=&k_\mu[\kslash,\gamma_\nu]+k_\nu[\gamma_\mu,\kslash].
\end{eqnarray}
In deriving the above result, we used identities such as
\begin{eqnarray}
\gamma_\mu\kslash\gamma_\nu-\gamma_\nu\kslash\gamma_\mu
&=&-\frac{1}{2}(\kslash[\gamma_\mu,\gamma_\nu]+[\gamma_\mu,\gamma_\nu]\kslash),
\label{eq_identity1}%
\\
\pslash\kslash\gamma_\nu-\gamma_\nu\kslash\pslash
&=&+\frac{1}{2}(\pslash[\kslash,\gamma_\nu]+[\kslash,\gamma_\nu]\pslash).
\label{eq_identity2}%
\end{eqnarray}

Now we manipulate $(1e)$. Taking the derivative, plugging in the
propagator $P^{\rho\sigma}$ and doing antisymmetrization, one
obtains
\begin{eqnarray}
\big[\partial_\nu^q(1e)_0\big]&=&-\big(K_{\mu\nu}^0+K_{\mu\nu}^1\big)P_L\frac{1}{2P^2Q_1}%
+E_{\mu\nu}P_L\frac{\delta_\xi}{4P^2Q_1Q_2},
\end{eqnarray}
where, using $p^2=m_i^2$ and the identity
\begin{eqnarray}
\kslash[\gamma_\mu,\gamma_\nu]\kslash
&=&k^2[\gamma_\mu,\gamma_\nu]-2([\gamma_\mu,\kslash]k_\nu+[\kslash,\gamma_\nu]k_\mu),
\label{eq_identity3}%
\end{eqnarray}
the second term is recast as follows:
\begin{eqnarray}
E_{\mu\nu}&=&\kslash\big((\kslash+\pslash)[\gamma_\mu,\gamma_\nu]
+[\gamma_\mu,\gamma_\nu](\kslash+\pslash)\big)\kslash
\nonumber%
\\
&=&[(k+p)^2-m_i^2]K_{\mu\nu}^1-k^2K_{\mu\nu}^0
\nonumber%
\\
&&+2\big\{k_\mu(\pslash[\kslash,\gamma_\nu]+[\kslash,\gamma_\nu]\pslash)
+k_\nu(\pslash[\gamma_\mu,\kslash]+[\gamma_\mu,\kslash]\pslash)\big\}.
\end{eqnarray}
Application of the limiting EoMs gives finally
\begin{eqnarray}
\big[\partial_\nu^q(1e)_0\big]&\leftrightharpoons&-\big(K^1_{\mu\nu}
+2m_i[\gamma_\mu,\gamma_\nu]\big)\frac{1}{4P^2Q_1}%
\nonumber%
\\
&&+\big(PK^1_{\mu\nu}+(m_\alpha^2-m_i^2)K^1_{\mu\nu}
-2m_ik^2[\gamma_\mu,\gamma_\nu]+4m_iK^2_{\mu\nu}\big)\frac{\delta_\xi}{8P^2Q_1Q_2}.
\label{eq_dipole1e}%
\end{eqnarray}

The graph Fig. 1(a) involves the triple gauge coupling and double
gauge boson propagators, making it the most complicated to evaluate.
We outline how this is accomplished. Taking the derivative and doing
antisymmetrization we have
\begin{eqnarray}
[\partial_\nu^q(1a)_0]&=&+P^{-1}\gamma^\sigma(\kslash+\pslash)\gamma^\rho
P_LA_{\rho\sigma;\mu\nu},
\end{eqnarray}
where
\begin{eqnarray}
A_{\rho\sigma;\mu\nu}&=&-\frac{1}{2}\delta_\xi
k^2\bigg(\frac{G_{\mu\nu,\rho\sigma}^0}{Q_1^2Q_2}
+\frac{G_{\mu\nu,\rho\sigma}^2}{Q_1^2Q_2^2}\bigg)
-\delta_\xi\frac{G_{\mu\nu,\rho\sigma}^2}{Q_1^2Q_2}
-\frac{3}{2}\frac{G_{\mu\nu,\rho\sigma}^0}{Q_1^2},
\end{eqnarray}
and
\begin{eqnarray}
G_{\mu\nu,\rho\sigma}^0&=&g_{\nu\sigma}g_{\mu\rho}-g_{\nu\rho}g_{\mu\sigma},
\\
G_{\mu\nu,\rho\sigma}^2&=&k_\mu(g_{\nu\rho}k_\sigma-g_{\nu\sigma}k_\rho)
-k_\nu(g_{\mu\rho}k_\sigma-g_{\mu\sigma}k_\rho).
\end{eqnarray}
The contraction with $G^0$ is standardized using Eq.
(\ref{eq_identity1}) into $(K^0_{\mu\nu}+K_{\mu\nu}^1)P_L$, while
the contraction with $G^2$ yields, by making use of Eq.
(\ref{eq_identity2}),
\begin{eqnarray}
\big(k_\mu(\pslash[\kslash,\gamma_\nu]+[\kslash,\gamma_\nu]\pslash)
+k_\nu(\pslash[\gamma_\mu,\kslash]+[\gamma_\mu,\kslash]\pslash)\big)P_L,
\end{eqnarray}
which reduces to $m_iK_{\mu\nu}^2$ using EoMs. The final form is
\begin{eqnarray}
[\partial_\nu^q(1a)_0]&\leftrightharpoons&-\big(K^1_{\mu\nu}
+2m_i[\gamma_\mu,\gamma_\nu]\big) \bigg(\frac{\delta_\xi
k^2}{8PQ_1^2Q_2}+\frac{3}{8PQ_1^2}\bigg)
\nonumber%
\\
&& +2m_iK_{\mu\nu}^2\bigg(\frac{\delta_\xi k^2}{8PQ_1^2Q_2^2}
+\frac{\delta_\xi}{4PQ_1^2Q_2}\bigg).
\label{eq_dipole1a}%
\end{eqnarray}

To summarize our calculation thus far, the terms relevant to the
neutrino magnetic moment are given in Eqs.
(\ref{eq_dipole1a},\ref{eq_dipole1cd},\ref{eq_dipole1e},\ref{eq_dipole1f}).
The next task is to demonstrate the $\xi_W$ cancellation among those
terms. We first decompose $\delta_\xi k^2=Q_2-\xi_WQ_1$ to remove
$k^2$ from numerators in Eqs. (\ref{eq_dipole1a},\ref{eq_dipole1e}).
The $K_{\mu\nu}^2$ terms sum to
\begin{eqnarray*}
\frac{1}{2}m_i\delta_\xi K_{\mu\nu}^2\bigg(\frac{1}{PQ_1^2Q_2}
+\frac{1}{PQ_1Q_2^2}+\frac{1}{P^2Q_1Q_2}\bigg).
\end{eqnarray*}
For any of the three terms in the above, the $k_\mu$ and $k_\nu$
factors in $K_{\mu\nu}^2$ may be simultaneously replaced by
$(k+p)_\mu$ and $(k+p)_\nu$, because the resulted additional terms,
upon the loop integration, will be proportional to
\begin{eqnarray*}
p_\mu[\pslash,\gamma_\nu]+p_\nu[\gamma_\mu,\pslash],
\end{eqnarray*}
which vanishes when sandwiched between the initial and final
spinors. We make this replacement for the last term in the sum.
Using again $\partial_\mu Q_j^{-1}=-2k_\mu Q_j^{-2}$ and
$\partial_\mu P^{-1}=-2(k+p)_\mu P^{-2}$, the sum becomes
\begin{eqnarray}
&&-\frac{1}{4}m_i\delta_\xi\big([\kslash,\gamma_\nu]\partial_\mu
+[\gamma_\mu,\kslash]\partial_\nu\big)(PQ_1Q_2)^{-1}
\nonumber%
\\
&=&-\frac{1}{4}m_i\delta_\xi\Big[\partial_\mu\big([\kslash,\gamma_\nu](PQ_1Q_2)^{-1}\big)
-(\mu\leftrightarrow\nu)\Big]
+\frac{1}{2}m_i\delta_\xi[\gamma_\mu,\gamma_\nu](PQ_1Q_2)^{-1}.
\label{eq_K2}
\end{eqnarray}
The apparently $\xi_W$-dependent terms in the sum
$\displaystyle\sum_{x=a}^f[\partial_\nu^q(1x)_0]$, including the one
in Eq. (\ref{eq_K2}) but dropping total derivatives, are collected
below:
\begin{eqnarray}
\sum_{x=a}^f[\partial_\nu^q(1x)_0]_\xi&\leftrightharpoons&
m_i[\gamma_\mu,\gamma_\nu]\frac{\delta_\xi}{4PQ_1Q_2}
\nonumber%
\\
&&+\frac{m_i}{m_W^2}\frac{1}{4P^2Q_2}
\big(K^1_{\mu\nu}m_i+[\gamma_\mu,\gamma_\nu](m_i^2-m_\alpha^2+\xi_Wm_W^2)\big).
\label{eq_dipolexi}%
\end{eqnarray}
The integral of the above second term is simplified using Eq.
(\ref{eq_integral1}) and EoMs, while the first one is split by
$\delta_\xi(Q_1Q_2)^{-1}=m_W^{-2}(Q_1^{-1}-Q_2^{-1})$, so that the
$\xi_W$ dependence disappears completely from the sum:
\begin{eqnarray}
\sum_{x=a}^f[\partial_\nu^q(1x)_0]_\xi&\leftrightharpoons&
m_i[\gamma_\mu,\gamma_\nu]\frac{1}{4m_W^2}\bigg(\frac{1}{PQ_1}-\frac{1}{P^2}\bigg).
\label{eq_dipolexi0}
\end{eqnarray}
Adding the above with the terms that are explicitly
$\xi_W$-independent, we obtain the final sum of terms contributing
to the neutrino magnetic moment:
\begin{eqnarray}
\sum_{x=a}^f[\partial_\nu^q(1x)_0]&\leftrightharpoons&
m_i[\gamma_\mu,\gamma_\nu]\frac{1}{4m_W^2}\bigg(\frac{1}{PQ_1}-\frac{1}{P^2}\bigg)
-\big(K^1_{\mu\nu}+2m_i[\gamma_\mu,\gamma_\nu]\big)\frac{1}{2PQ_1^2}
\nonumber%
\\
&&+\big([m_W^{-2}(m_\alpha^2-m_i^2)-2]K^1_{\mu\nu}
-6m_i[\gamma_\mu,\gamma_\nu]\big)\frac{1}{8P^2Q_1}.
\label{eq_sum}%
\end{eqnarray}

From Eqs. (\ref{eq_A},\ref{eq_Gamma},\ref{eq_sum}) and the loop
integrals defined in the appendix, we obtain for the neutrino
$\nu_i$ the magnetic form factor at the vanishing momentum transfer,
\begin{eqnarray}
F_2(0)&=&-\frac{g_2^2}{(4\pi)^2}\frac{2m_i^2}{m_W^2}
\sum_\alpha|V_{\alpha i}|^2\bigg[%
\frac{1}{4}I_1+J_2 -\frac{1}{2}K_2
+\frac{3}{4}J_1-\frac{1}{8}(2-x_\alpha+y_i)K_1\bigg],
\label{eq_F20}%
\end{eqnarray}
where $I_1,~J_{1,2},~K_{1,2}$ are functions of the mass ratios
$x_\alpha=m_\alpha^2/m_W^2$ and $y_i=m_i^2/m_W^2$. This result is
indeed manifestly gauge independent in the class of $R_\xi$ gauges.

\section{Evaluation in unitary gauge}
\label{sec:Ugau}

Working in unitary gauge means that the limit $\xi_W\to\infty$ is
taken before the loop integrals are evaluated. Since $\xi_W$ appears
exclusively in the propagators of the $W^\pm$ gauge bosons, would-be
Goldstone bosons $G^\pm$ and the ghosts $c^\pm$, only the gauge
boson propagator survives the limit,
\begin{eqnarray}
P_{\mu\nu}(k)\to\bar P_{\mu\nu}(k)=(g_{\mu\nu}-m_W^{-2}k_\mu k_\nu)
Q_1^{-1},
\end{eqnarray}
and thus only the pure-$W^\pm$ graphs $(a,~e)$ in Fig. 1 and
$(a,~c)$ in Fig. 2 remain. We have presented our calculation in
$R_\xi$ gauge in a way that can be easily adapted for unitary gauge.

For the charge contribution from Fig. 1 we take the limit
$\xi_W\to\infty$ in the integrand (\ref{eq_1aplus1e0}) where only
the total derivative term survives:
\begin{eqnarray}
[(1a)+(1e)]_0&\to&\partial_\mu\big\{\gamma_\sigma(\kslash+\pslash)\gamma_\rho
P_LP^{-1}\bar P^{\rho\sigma}\big\},
\end{eqnarray}
whose integral vanishes in dimensional regularization. The relevant
terms from the photon-$Z$ mixing energy are obtained from Eqs.
(\ref{eq_2a},\ref{eq_2c}), or more readily from Eq.(\ref{eq_sum2x}),
\begin{eqnarray}
[(2a)+(2c)]&\to&\partial_\mu\big\{-2c_W^2(d-1)(k_\nu
Q_1^{-1})\big\},
\end{eqnarray}
whose integral again vanishes. Thus the vanishing of charge at one
loop occurs in unitary gauge in a stronger manner: each of the
contributions from the proper vertices and the mixing energy
vanishes separately.

The magnetic form factor $F_2(0)$ can also be obtained from
intermediate steps in subsec \ref{subsec:mag}. We can sum Eqs.
(\ref{eq_dipole1a},\ref{eq_dipole1e}) and take the limit
$\xi_W\to\infty$, or cope directly with the total of all graphs
since we know only Figs. 1$(a,e)$ survive the limit. The latter
point can also be seen from explicit results in
Eqs.(\ref{eq_dipole1cd},\ref{eq_dipole1f}). Dropping the total
derivatives and sending $\xi_W\to\infty$, the potentially
$\xi_W$-dependent part of the total in $R_\xi$ gauge, Eq.
(\ref{eq_dipolexi}), goes exactly to Eq. (\ref{eq_dipolexi0})
without additional manipulations. The result in Eq. (\ref{eq_F20})
is thus recovered in unitary gauge.

\section{Summary}
\label{sec:summary}

The electromagnetic properties of neutrinos are an interesting topic
that is potentially relevant to various astrophysical phenomena and
laboratory measurements. Although we know from principles that the
charge and dipole moments of a Dirac neutrino are physical
quantities and cannot depend on computational methods or the choice
of gauge in a consistent theory, this has never been explicitly
examined before in a satisfactory manner even at one loop. We have
studied this issue in the minimally extended standard model that
incorporates neutrinos masses and mixing. We demonstrated at one
loop in both $R_\xi$ and unitary gauges that the magnetic moment and
vanishing charge are indeed gauge-independent quantities. This
statement is exact in the sense that it is true for any values of
various masses and the lepton mixing matrix as long as the latter is
unitary. We have accomplished this by manipulating directly the
integrands of loop integrals and employing simple algebraic
identities like
(\ref{eq_identity1},\ref{eq_identity2},\ref{eq_identity3}) and
integral relations like (\ref{eq_integral1}). We believe this
approach is advantageous over the one that handles the results of
loop integration, and may be useful in other contexts. Finally, we
mention that various approximations to our exact one-loop result for
the magnetic moment in Eq. (\ref{eq_F20}) are possible. For
instance, when all neutrinos and charged leptons are much lighter
than the weak gauge bosons as is the case in SM, we have from the
explicit results in the appendix that $F_2(0)\approx
-(3G_Fm_i^2)/(4\pi^2\sqrt{2})$, where the mixing matrix drops out
from the leading term, so that the interaction potential of the
neutrino $\nu_i$ of mass $m_i$ and spin $\vec S$ with an external
magnetic moment is, $V\approx(3eG_Fm_i)/(4\pi^2\sqrt{2})\vec
S\cdot\vec B$, recovering the well-known result in the literature.

\vspace{0.5cm}
\noindent %
{\bf Acknowledgement}

WTH would like to thank the members of theory group at Nankai
University for hospitality during a long-term visit when this work
was conducted. This work was supported in part by the grant
NSFC-11025525 and by the Fundamental Research Funds for the Central
Universities No.65030021.

\vspace{0.5cm}
\noindent %
{\Large\bf Appendix: some useful integrals}
\vspace{0.5cm}%

We list some loop integrals relevant to our evaluation of the
magnetic moment. The following relation is used in sec
\ref{sec:Rgau} for reduction of terms:
\begin{eqnarray}
2p^2\int_k\frac{k_\alpha}{D_1^{n_1}D_2^{n_2}}
&=&p_\alpha\int_k\bigg(\frac{1}{D_1^{n_1-1}D_2^{n_2}}
-\frac{1}{D_1^{n_1}D_2^{n_2-1}}+\frac{m_1^2-m_2^2-p^2}{D_1^{n_1}D_2^{n_2}}\bigg),
\label{eq_integral1}%
\end{eqnarray}
where $D_1=(k+p)^2-m_1^2,~D_2=k^2-m_2^2$. Using notations in Eq.
(\ref{eq_prop}) with $p^2=m_i^2$, the basic integral is
\begin{eqnarray}
&&\int_k\bigg(\frac{1}{Q_1^2}-\frac{1}{PQ_1}\bigg)=\frac{i}{(4\pi)^2}I(x_\alpha,y_i),
\end{eqnarray}
where $x_\alpha=m_\alpha^2/m_W^2$ and $y_i=m_i^2/m_W^2$. For
simplicity, we also define the integrals
\begin{eqnarray}
&&\int_k\bigg(\frac{1}{PQ_1}-\frac{1}{P^2}\bigg)=\frac{i}{(4\pi)^2}I_1(x_\alpha,y_i),
\\
&&\int_k\frac{1}{P^2Q_1}=-\frac{i}{(4\pi)^2}\frac{1}{m_W^2}J_1(x_\alpha,y_i),
\\
&&\int_k\frac{1}{PQ_1^2}=-\frac{i}{(4\pi)^2}\frac{1}{m_W^2}J_2(x_\alpha,y_i),
\\
&&\int_k\frac{2k_\mu}{P^2Q_1}=\frac{i}{(4\pi)^2}\frac{p_\mu}{m_W^2}K_1(x_\alpha,y_i),
\\
&&\int_k\frac{2k_\mu}{PQ_1^2}=\frac{i}{(4\pi)^2}\frac{p_\mu}{m_W^2}K_2(x_\alpha,y_i).
\end{eqnarray}
The parametric integral for $I(s,t)$ is
\begin{eqnarray}
I(x,y)=\int_0^1dt~\ln\big[xt+(1-t)-yt(1-t)-i0^+\big].
\end{eqnarray}
The other functions are related to it by
\begin{eqnarray}
I_1(x,y)&=&\ln x-I(x,y),
\\
J_1(x,y)&=&\frac{\partial}{\partial x}I(x,y),
\\
J_2(x,y)&=&J_1(1/x,y/x),
\\
K_1(x,y)&=&y^{-1}[I_1(x,y)+(1+y-x)J_1(x,y)],
\\
K_2(x,y)&=&y^{-1}[I(x,y)+(1+y-x)J_2(x,y)].
\end{eqnarray}
Note that the singularity at $y=0$ is spurious since the original
integrals are smooth there.

The analytic result for $I$ is known for all parameter regions, but
we only record it for the case relevant to SM, i.e., for $0\le
y<x\ll 1$,
\begin{eqnarray}
I(x,y)&=&-2-\frac{1}{2y}(1-x-y)\ln x +\frac{\lambda}{2y}\ln R,
\end{eqnarray}
where
\begin{eqnarray}
\lambda=(1+x^2+y^2-2x-2y-2xy)^{1/2},~R=\frac{1+x-y-\lambda}{1+x-y+\lambda}.
\end{eqnarray}
The other two functions are
\begin{eqnarray}
J_1(x,y)&=&-\frac{1-x+y}{2y\lambda}\ln R +\frac{1}{2y}\ln x,
\\
J_2(x,y)&=&-J_1(x,y)-\frac{1}{\lambda}\ln R.
\end{eqnarray}

\noindent %

\end{document}